\documentclass{ifacconf}
\usepackage[utf8]{inputenc}

\usepackage{graphicx}      % include this line if your document contains figures
\usepackage{natbib}        % required for bibliography
\usepackage{siunitx}
\usepackage{amsmath}
\usepackage{amsfonts}
\usepackage{url}

\DeclareMathOperator*{\argmin}{arg\,min}

\usepackage{makecell} % for the text spanning two rows in a table cell

\usepackage{subcaption}
\usepackage{todonotes}

\begin{document}
\begin{frontmatter}

\title{Optimization-based Feedback Manipulation Through an Array of Ultrasonic Transducers}

\author{Josef Matouš\quad} 
\author{\quad Adam Kollarčík\quad}
\author{Martin Gurtner\quad}
\author{Tomáš Michálek\quad}
\author{Zdeněk Hurák}

\address{Faculty of Electrical Engineering\\Czech Technical University in Prague\\Technicka 2, 166 27 Praha 6, Czech Republic\\\{josef.matous,adam.kollarcik\}@fel.cvut.cz}

\begin{abstract}                % Abstract of not more than 250 words.
In this paper we document a novel laboratory experimental platform for non-contact planar manipulation (positioning) of millimeter-scale objects using acoustic pressure.
The manipulated objects are either floating on a water surface or rolling on a solid surface.
The pressure field is shaped in real time through an 8-by-8 array (matrix) of ultrasonic transducers.
The transducers are driven with square voltages whose phase-shifts are updated periodically every 20 milliseconds based on the difference between the desired and true (estimated from video) position.
Numerical optimization is used within every period of a discrete-time feedback loop to determine the phase shifts for the voltages.
The platform can be used as an affordable testbed for algorithms for non-contact manipulation through arrays of actuators as all the design and implementation details for the presented platform are shared with the public through a dedicated git repository.
The platform can certainly be extended towards higher numbers of simultaneously yet independently manipulated objects and larger manipulation areas by expanding the transducer array.  
\end{abstract}

\begin{keyword}
Distributed manipulation, non-contact manipulation, optimization-based control, acoustic manipulation, acoustophoresis, distributed control
\end{keyword}

\end{frontmatter}
%===============================================================================

\section{Introduction}
Methods of high-accuracy manipulation (positioning or orientation or both) of objects using conventional mechanical manipulators with grippers, needles or tweezers as their end effectors face challenges when: 1) the manipulated objects are fragile, 2) the manipulated objects are very small or 3) there is not just one but several (possibly many) objects to be manipulated simultaneously and independently.
Prominent examples of such situations are analysis of biological objects such as cells and assembly processes for technological components.
Principles of non-contact manipulation are appealing in these situations.
Among the most studied and used are laser tweezers, controlled pressure-driven flow, several phenomena derived from electric field such as dielectrophoresis, electrophoresis, electrorotation or electroosmosis,  and magnetic manipulation, including magnetophoresis.
An overview of these (as well as some more explanation of the motivation) is given in \cite{zemanek_thesis_2018}.
In this paper, we present yet another physical phenomenon that was fully harnessed for feedback manipulation---non-contact planar manipulation by shaping a pressure field through an array of ultrasonic transducers.
The idea is not new, as we carefully document in the next section, but we developed a novel experimental platform that might be interesting and useful as a testbed for research in a broader area of non-contact manipulation using arrays of actuators/generators.

\section{Related work using ultrasonic transducers}
There are numerous applications using an array of ultrasonic transducers. For instance,~\cite{marzo_ghost_2015} used an array of ultrasonic transducers to transform a sand layer or a surface of a liquid into an interactive canvas.
The transducers are driven by voltage signals with appropriate phase shifts to generate acoustic pressure focal points in desired positions.
Controlled positioning of these focal points enables drawing in the sand, inducing flow in a fluid, and to create, move with, or pop bubbles in a soap solution.
In~\cite{marzo_holographic_2015}, a similar platform was used as an ultrasonic levitator. Instead of the focal points, acoustic traps are generated, which are capable of supporting small and light objects (\emph{e.g.}, Styrofoam particles or small droplets of liquid) in midair.
The levitated objects can then be moved by gradual repositioning of the traps themselves.
Both of these platforms, and majority of others (\cite{long_rendering_2014,Ochiai2014Pixie,courtney_independent_2014}), are based on open-loop control systems; they do not measure position of the manipulated objects.
In contrast, \cite{marshall_ultra-tangibles:_2012} developed a so-called ``Ultra-Tangibles'' platform that utilizes a closed-loop control system. This platform is designed for interactive manipulation with multiple objects using ultrasound air-pressure waves. For their generation, they used four rectangular arrays of ultrasonic transducers (two $15\times3$ arrays and two $9\times3$ arrays) arranged in a rectangle around the manipulation area. Individual signals driving the transducers are generated by separate microcontrollers (\emph{i.e.}, there is one microcontroller per transducer). Therefore, this platform is rather complicated and difficult to reproduce. In contrast, the platform that we designed and described in this paper is relatively easy to reproduce and simple to program. It is also suitable for rapid prototyping (reprogramming).

\section{Platform description}
\begin{figure}[b]
	\centering
	\includegraphics[width=\columnwidth]{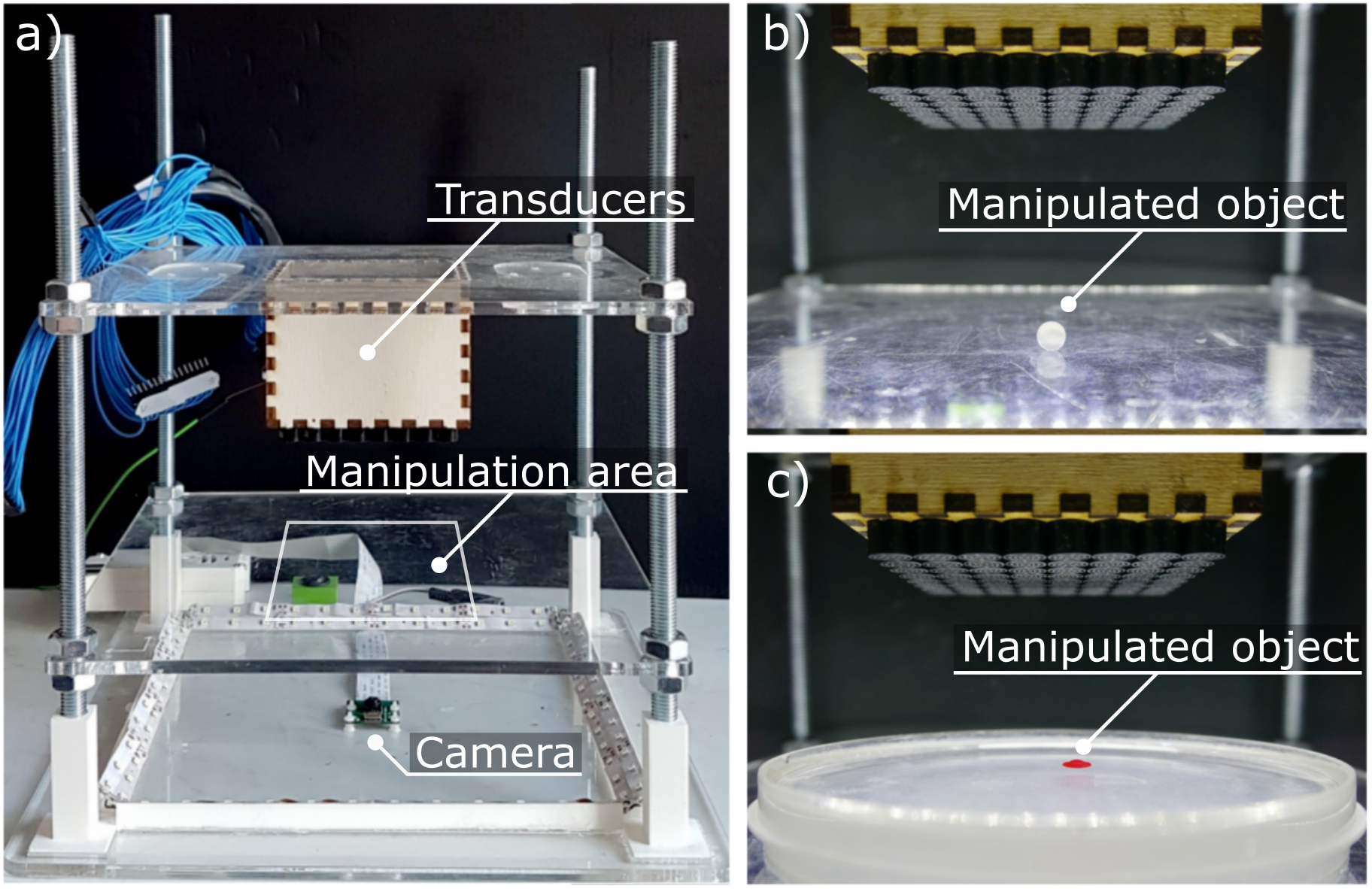}
	\caption{A photo of the platform as a whole in a), when an object on a solid surface is manipulated in b) and when an object floating in a shallow pool of water is manipulated in c).}
	\label{fig:platformPhoto}
\end{figure}

The photos of the platform are in Fig.~\ref{fig:platformPhoto}. Proceeding from the top to the bottom, there is an array of $8\times8$ \textit{MURATA MA40S4S} ultrasonic transducers mounted to the top plate, a manipulated object placed at the middle plate and light bars and a camera capturing the manipulation area from bottom mounted at the bottom plate.
The manipulated object can be either floating in a shallow pool of water (as shown in Fig. 1b) or rolling (in case of a spherical shape) on a solid flat surface (as shown in Fig. 1c).
All parts of the platform are designed to be easily reproducible by anyone having an access to a laser cutter and a 3D printer.
In addition to what can be seen in Fig.~\ref{fig:platformPhoto}, the complete platform also consists of a \textit{Raspberry Pi} computer, a driver (electrical circuit) for the transducers, and a power supply unit for the whole platform. These parts are described in more detail in the following paragraphs.

\begin{figure}[t]
	\centering
	\includegraphics[width=0.9\columnwidth]{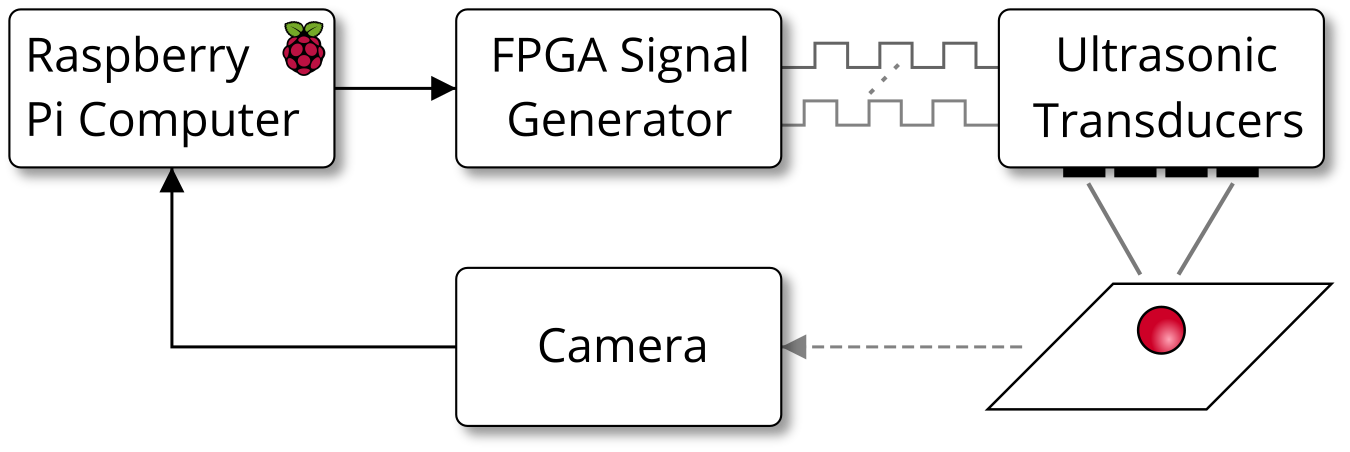}
	\caption{A principal signal diagram of the whole platform.}
	\label{fig:platformSigDiag}
\end{figure}

A signal diagram of the whole platform is in Fig.~\ref{fig:platformSigDiag}. 
The ultrasonic transducers are driven by a custom-made signal generator based on an easily accessible FPGA development board \textit{DE0-nano} by \textit{Terasic} and a custom-made shield (add-on).
The FPGA board generates 64 phase-shifted square waves where the phase shift can be set in multiples of $\pi/180$ every few milliseconds via USB from the Raspberry Pi.
The shield amplifies the \SI{3.3}{\volt} square waves generated by the FPGA to \SI{16}{\volt} so that they can drive the transducers.

The Raspberry Pi is the brain of the whole platform.
It runs the control algorithm, it estimates the position of the manipulated object by processing the images of the manipulation area captured by \textit{Raspberry Pi Camera Module V1.3}, and it also communicates with the signal (voltage) generator (via USB).
We justify our choice of Raspberry Pi by its accessibility, wide user base and good support in \textit{Matlab} and \textit{Simulink}, which enables rapid prototyping of control algorithms.

Technical drawings of all the parts, PCB designs and all the software is freely available at the platform's GitHub repository\footnote{\url{https://github.com/aa4cc/AcouMan}}.

\section{Modeling}
In this section, we describe a mathematical model of the platform, which is accurate enough yet still sufficiently simple for control purposes.
It consists of two coupled but separately modeled subsystems.
The first one is responsible for the calculation of the acoustic pressure field as a function of the transducers' phase delays, and the second one predicts the motion of the object of interest as an effect of the related acoustic force.

\subsection{Modeling the transducer}

We start with modeling the field of a single transducer.
Then, assuming the principle of superposition, we model the whole array by summing the contributions from individual transducers.

Complex acoustic pressure of the $i^{\rm th}$ transducer, $p_i$, depends both on the position, and on the phase delay.
However, the formula can be split to position- and delay-dependent terms:
\begin{equation}
	p_i(\mathbf{x},\varphi_i) = M_i(\mathbf{x}) \mathrm{e}^{j\varphi_i},
\end{equation}
where $\mathbf{x}=[x,y,z]$ is the position of the evaluated point, $\varphi_i$ is the phase delay, $j$ is the imaginary unit, and $M_i$ is the complex acoustic pressure for zero phase. 

\begin{figure}[h]
	\centering
	\includegraphics[width = 0.16\textwidth]{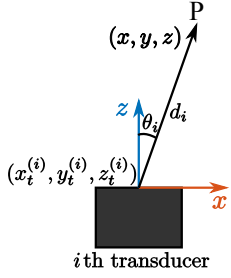}
	\caption{A sketch of the distance and the angle from the $i^\mathrm{th}$ transducer.}
	\label{fig:transducerModel}
\end{figure}

In literature, there are numerous methods modeling the pressure $M_i$ (\cite{marzo_holographic_2015,courtney_dexterous_2013,long_rendering_2014}).
We use the following model proposed in \cite{marzo_ghost_2015}:
\begin{equation}
	M_i = Af_{\mathrm{dir}}(\theta_i)\frac{1}{d_i}\mathrm{e}^{jkd_i},
\label{eq:ghostTouchModel}
\end{equation}
where $A$ is the power of the transducer\footnote{The so-called ``power of the transducer'' is actually the acoustic pressure generated by the transducer at a given distance, \emph{e.g.} one meter.
Therefore, the unit of this quantity is \textit{Pascal meter} (\si{\pascal\metre}).}, $f_{\mathrm{dir}}$ is the so-called directivity function representing the polar pattern of the transducer, $k$ is the wavenumber, $d_i$ is the distance from the $i^\mathrm{th}$ transducer, and $\theta_i$ is the angle between the axis of the transducer and a line connecting it with the point $\mathbf{x}$.
The distance and the angle are shown in Fig.~\ref{fig:transducerModel}.

For the directivity function, we employ a far-field model of a vibrating circular piston source~\cite[p.~179-182]{kinsler_fundamentals_2000}:
\begin{equation}
	f_{\mathrm{dir}} = \frac{2{\rm J}_1(kr\sin\theta_i)}{kr\sin\theta_i},
\end{equation}
where ${\rm J}_1$ is the first-order Bessel function of the first kind, and $r$ is the known radius of the transducer.

\begin{figure}[b]
	\centering
	\includegraphics[width = .45 \textwidth]{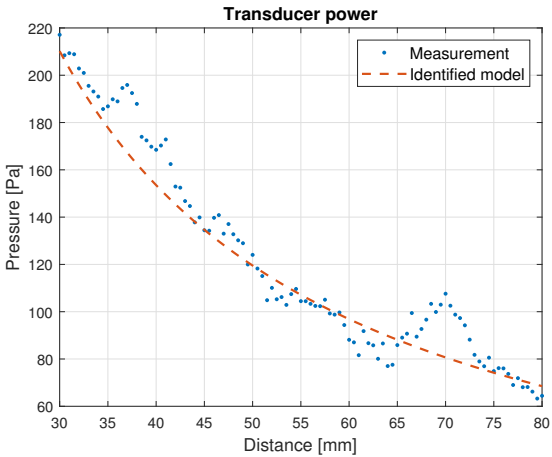}
	\caption{Acoustic pressure generated by a single transducer with respect to the distance from its surface.}
	\label{fig:meas_trans_power}
\end{figure}

In order to determine the power of the transducer, we measured the pressure generated at distances from \SIrange{30}{80}{\milli\metre} using the \textit{GRAS Type40DP} polarized microphone mounted on a motorized stage.
Interpolating the measured data by a rational function, we identified the transducer power, $A$ from~(\ref{eq:ghostTouchModel}), to be approximately \SI{6.8}{\pascal\metre}.
The comparison of the measurement and the model of a single transducer is shown in Fig.~\ref{fig:meas_trans_power}.

\subsection{Modeling the pressure field created by a transducer array}

As mentioned above, we assume the principle of superposition.
Therefore, the total acoustic pressure at a given point $\mathbf{x}$ with a given vector of phase delays $\boldsymbol{\Phi} = \left[ \varphi_1\,,\,\varphi_2\,,\,\ldots\,,\,\varphi_N \right]$, where $N$ is the number of transducers, is:
\begin{equation}
    \label{eq:pressureModel}
	p(\mathbf{x},\boldsymbol{\Phi}) = \sum_{i=1}^{N}p_i(\mathbf{x},\varphi_i).
\end{equation}

\begin{figure}[h]
	\centering
	\includegraphics[width = \columnwidth]{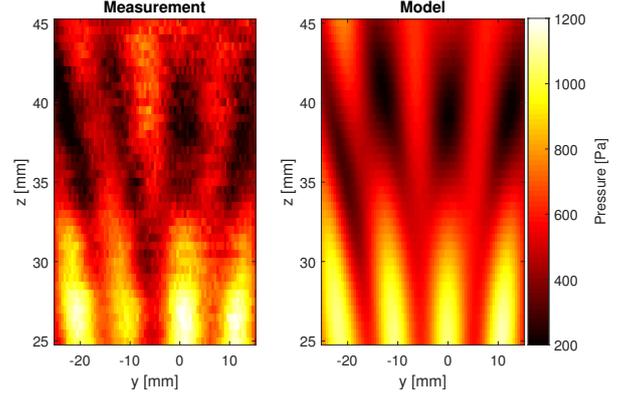}
	\caption{Acoustic pressure field in $y$-$z$ plane ($x=0$) of the array. All transducers have the same phase.}
	\label{fig:meas_field}
\end{figure}

To verify the complete model of the array, we scanned the array in $y$-$z$ plane with \SI{0.5}{\milli\metre} resolution.
The comparison of the measurement and the model is shown in Fig. \ref{fig:meas_field}.

\subsection{Modeling the motion of the manipulated object}
\label{subsec:object_model}

Although there exist analytical physics-based models of the acoustic pressure radiation forces (see \cite{king_acoustic_1934,bruus_acoustofluidics_2012-1}), the resulting computations are too complex to run several times per second.
Therefore, we use an approximate model in the form of a second-order linear system identified from experiments.

We assume that the force exerted on the object is proportional to the negative gradient of the acoustic pressure amplitude at a given distance $R$ from the object's position (see Fig.~\ref{fig:press2force}).
In addition, we assume the drag and friction forces to be proportional to velocity.
Then, the equation of motion of the manipulated object along one axis is:
\begin{equation}
	m\ddot{x} = c_\mathrm{f}\dot{x} + c_\mathrm{p}\left(|p| - P_{\rm off} \right),
\label{eq:modelNewton}
\end{equation}
where $x$ is the position of the object, $m$ is its mass, $c_\mathrm{f}$ is the coefficient of friction, $c_\mathrm{p}$ is the acoustic pressure to force conversion constant, $|p|$ is the modulus of the high-pressure point, and $P_{\rm off}$ is an offset related to the ambient pressure (\emph{i.e.}, an acoustic pressure generated outside the high-pressure point).

Using $\Delta p = |p| - P_{\rm off}$ as the control input, model \eqref{eq:modelNewton} becomes linear and the unknown coefficients can be obtained by various methods for identification of linear systems.

\section{Control architecture}
\begin{figure}[t]
    \centering
    \includegraphics[width = \columnwidth]{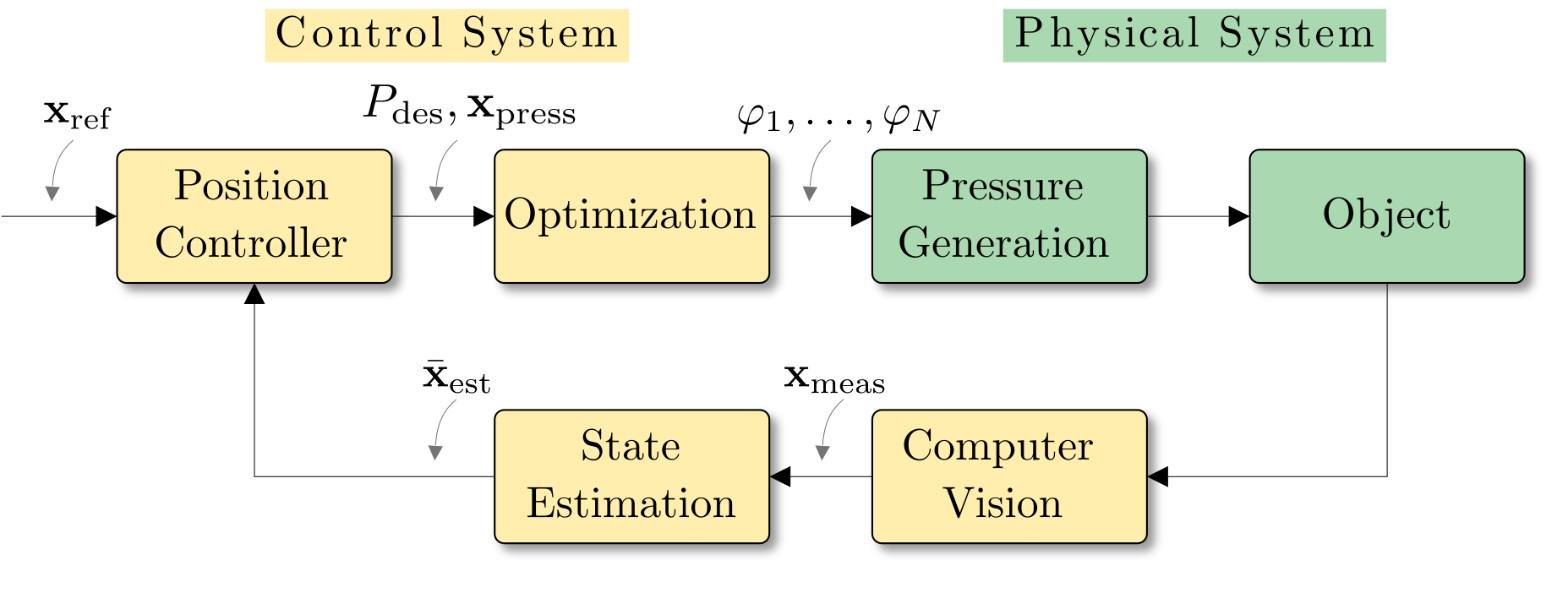}
    \caption{Signal flow of the control system.}
    \label{fig:controlScheme}
\end{figure}

We first give a concise overall description of the architecture of the control system and then we describe the individual parts in some more detail.
A signal diagram of the feedback control system is sketched in Fig.~\ref{fig:controlScheme}.
Both the reference (desired) position $\mathbf{x}_\mathrm{ref}=[x_\mathrm{ref}, y_\mathrm{ref}]$ for the object and its estimated (from video) position are fed to the position controller (algorithms run on Raspberry Pi).
Based on the difference between the reference and true positions, the control algorithm computes the desired position $\mathbf{x}_\mathrm{press}$ of the local maximum of pressure and its amplitude $P_\mathrm{des}$, such the manipulated object feels force pushing it towards the reference position.
This information is then processed by an optimization algorithm that computes the phase shifts $\varphi_1,\dots,\varphi_N$ of square (voltage) signals driving the transducers.
These phase shifts are then sent to the signal generator.
After applying the voltages of the specified phase shifts, the local maximum  of pressure is generated and, consequently, the manipulated object changes its position.
The new position is sensed by a computer vision system and sent to a state estimator, which compensates for the measurement delay inevitably caused by processing the captured image.
This feedback loop works with the sampling frequency of $50\,\mathrm{Hz}$.

\subsection{Position Controller} % (fold)
\begin{figure}[t]
	\centering
	\includegraphics[width=0.5\columnwidth]{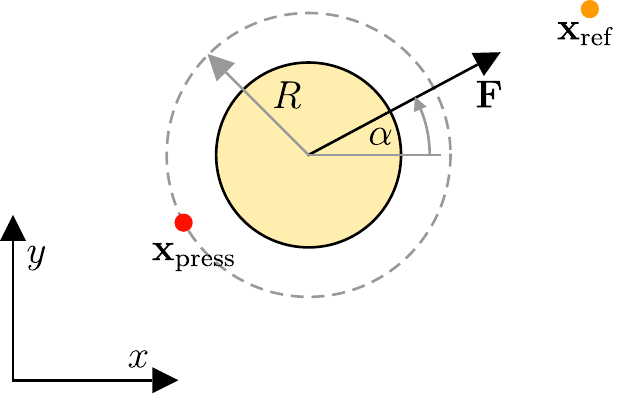}
	\caption{Calculation of the position of the high pressure point where yellow circle represents the manipulated object.}
	\label{fig:press2force}
\end{figure}

\label{sub:position_controller}
To steer the manipulated object towards the reference position, we need to determine the location and amplitude of the local maximum of the acoustic pressure field.
To do so, we evaluate the position regulation error, that is, the deviation of the estimated and the reference positions $[\delta_x, \delta_y]=[x_\mathrm{ref} - x_\mathrm{est}, y_\mathrm{ref} - y_\mathrm{est}]$.
Then, we feed $[\delta_x, \delta_y]$ separately to two simple proportional-integral-derivative (PID) controllers outputting the components of the force $\mathbf{F}=[F_x, F_y]$ that is needed to push the manipulated object towards the reference position.
The amplitude and position of the local maximum of pressure is then computed by
\begin{align}
	P_\mathrm{des} &= \frac{1}{c_\mathrm{p}}\|\mathbf{F}\|_2 + P_\mathrm{off}, \\
	x_\mathrm{press} &= x_\mathrm{est} - R\cos{\alpha}, \\
	y_\mathrm{press} &= y_\mathrm{est} - R\sin{\alpha},
\end{align}
where $R$ is the distance of the pressure point from the center of the object and $\alpha$ is the angle between the x-axis and the force vector $\mathbf{F}$, as shown in Fig.~\ref{fig:press2force}.
Constants $c_\mathrm{p}$ and $P_\mathrm{off}$ are the parameters of the linear model discussed in the previous section.
% subsection position_controller (end)

\subsection{Optimization} % (fold)
\label{sub:optimization}
The fundamental actuation principle is to create an area of high acoustic pressure, which forces the manipulated object to move to a lower pressure zone.
Having the model of the acoustic pressure~\eqref{eq:pressureModel}, we can optimize the phase shifts so that the desired high pressure zone is generated at a specified location.
This can be formulated as the following optimization problem
\begin{equation}
\label{eq:optimProblem}
    \boldsymbol{\Phi}^* = \argmin_{\boldsymbol{\Phi}\in\mathbb{R}^N} \| |p(\mathbf{x}_\mathrm{press}, \boldsymbol{\Phi})|^2 -  P_\mathrm{des}^2\|_2^2,
\end{equation}
where $P_\mathrm{des}$ is the desired value of the pressure at the desired point $\mathbf{x}_\mathrm{press}$\footnote{Interestingly, it can be shown that optimization problem~\eqref{eq:optimProblem} is equivalent to the problem emerging in feedback manipulation by dielectrophoresis where one also optimizes phase shifts of signals driving an array of actuators (for details, see \cite{zemanek_phase-shift_2018}).}.

There are two troubling aspects regarding the optimization problem~\eqref{eq:optimProblem}.
First, it can be shown that the problem~\eqref{eq:optimProblem} is non-convex.
It is therefore difficult to obtain a global minimum (of the difference between the desired and achieved pressures).
Second trouble is even more serious: the cost function does not perfectly express what we need---it does not capture the requirement that the pressure generated at the given location is a local maximum (of pressure).
Nevertheless, it turns out that when the optimization problem is initialized with random phase shifts and $P_\mathrm{des}$ is physically achievable, the L-BFGS solver we use for solving~\eqref{eq:optimProblem} almost always ends up with $\boldsymbol{\Phi}^*$ generating a local maximum of pressure.

For reader's convenience, we also show how the pressure and its gradient---which are needed for solving the optimization problem~\eqref{eq:optimProblem}---can be efficiently computed.
Let us define a vector $\mathbf{m}=[M_1(\mathbf{x}_\mathrm{press}),\dots,M_N(\mathbf{x}_\mathrm{press})]^\mathrm{T}$ and a vector $\mathbf{u}=[\mathrm{e}^{j\varphi_1}, \dots, \mathrm{e}^{j\varphi_N}]^\mathrm{T}$.
For brevity, we omit the dependence of $\mathbf{u}$ on $\boldsymbol{\Phi}$ and of $\mathbf{m}$ on $\mathbf{x}_\mathrm{press}$.
Now, the pressure~\eqref{eq:pressureModel} can expressed as \begin{equation}
\label{eq:acu_pressure_quad}
  |p|^2 = \mathbf{u}^\dagger \, \bar{\mathbf{m}}\mathbf{m}^\mathrm{T} \, \mathbf{u},
\end{equation}
where $\bar{\mathbf{m}}$ is a vector of complex conjugated components of $\mathbf{m}$ and $\mathbf{u}^\dagger$ denotes the conjugate transpose of $\mathbf{u}$.
To get rid of the complex numbers, we reformulate~\eqref{eq:acu_pressure_quad} to
\begin{equation}
\label{eq:acu_pressure_quad2}
  |p|^2 = \mathbf{c}_{\boldsymbol{\Phi}}^\mathrm{T} \, \mathbf{P}_1 \, \mathbf{c}_{\boldsymbol{\Phi}} + \mathbf{s}_{\boldsymbol{\Phi}}^\mathrm{T} \, \mathbf{P}_1 \, \mathbf{s}_{\boldsymbol{\Phi}} + \mathbf{c}_{\boldsymbol{\Phi}}^\mathrm{T} \, \mathbf{P}_2 \, \mathbf{s}_{\boldsymbol{\Phi}},
\end{equation}
where vectors $\mathbf{c}_{\boldsymbol{\Phi}}$ and $\mathbf{s}_{\boldsymbol{\Phi}}$ are cosines and sines of phase shifts in $\boldsymbol{\Phi}$, that is $\mathbf{c}_{\boldsymbol{\Phi}}=[\cos(\varphi_1), \dots, \cos(\varphi_n)]^\mathrm{T}$ and $\mathbf{s}_{\boldsymbol{\Phi}}=[\sin(\varphi_1), \dots, \sin(\varphi_n)]^\mathrm{T}$.
Matrices $\mathbf{P}_1$ and $\mathbf{P}_2$---with the aid of matrix $\mathbf{p}=[\mathbf{m}_\mathrm{r}, \mathbf{m}_\mathrm{i}]$ composed of vectors of real ($\mathbf{m}_\mathrm{r}$) and imaginary ($\mathbf{m}_\mathrm{i}$) parts of $\mathbf{m}$---are defined as follows
\begin{equation}
  \mathbf{P}_1 =
  \mathbf{p}\,\mathbf{p}^\mathrm{T},
  \quad
  \mathbf{P}_2 =
    \mathbf{p}
    \begin{bmatrix}
      0 & -2\\
      2 & 0
    \end{bmatrix}
    \mathbf{p}^\mathrm{T}.
\end{equation}

Finally, the differentiation of the pressure~\eqref{eq:acu_pressure_quad2} with respect to $\boldsymbol{\Phi}$ gives:
\begin{equation}
\begin{split}
	\!\!\!\nabla_{\boldsymbol{\Phi}} |p|^2  &=
	  2\left(\mathrm{diag}(\mathbf{c}_{\boldsymbol{\Phi}})  \mathbf{p}  \begin{bmatrix}
      0 & 1\\
      -1 & 0
    \end{bmatrix} -\mathrm{diag}(\mathbf{s}_{\boldsymbol{\Phi}})  \mathbf{p} \right)\mathbf{p}^\mathrm{T}  \mathbf{c}_{\boldsymbol{\Phi}} \\
	  &+ 2\left( \mathrm{diag}(\mathbf{c}_{\boldsymbol{\Phi}})  \mathbf{p} + \mathrm{diag}(\mathbf{s}_{\boldsymbol{\Phi}})  \mathbf{p}  \begin{bmatrix}
      0 & 1\\
      -1 & 0
    \end{bmatrix} \right) \mathbf{p}^\mathrm{T}  \mathbf{s}_{\boldsymbol{\Phi}},
\end{split}
\end{equation}
where $\mathrm{diag}(\mathbf{c}_{\boldsymbol{\Phi}})$ is a diagonal matrix with components of $\boldsymbol{\Phi}$ on the diagonal.
% subsection optimization (end)

\subsection{Computer Vision} % (fold)
\label{sub:computer_vision}
The position of the manipulated object is measured by a computer vision system.
Images captured by the camera module are processed by a Python script\footnote{\url{https://github.com/aa4cc/raspi-ballpos}} for position detection of colored spherical objects.
The image is downscaled, and an area of interest is chosen based on the weighted pixel sum of the RGB components.
Then, the weighted pixel sum is computed again for the area of interest in the original non-downscaled image, and the location of the object's center in the picture is obtained from the average position of pixels exceeding a threshold value.
For some colors, an additional operation such as morphological erosion is required to filter the weighted sum.
Last, 2D homography is applied to acquire the position in the platform coordinates. We estimate the measurement resolution to be around 1 \si{\milli\metre} for both coordinates, based on our testing experiments.
% subsection computer_vision (end)

\subsection{State Estimation} % (fold)
\label{sub:state_estimation}

The primary purpose of the used state estimator is to deal with the non-negligible time delay present in the system.
The processing of each camera frame takes around $60\,\mathrm{ms}$ and solving of the optimization problem together with communication delays take another $20\,\mathrm{ms}$, which in total corresponds to a duration of 4 control periods.
We use a Kalman filter with augmented state to predict the position of the object at the exact time instant when a next control action is about to be applied.
We discretized the model described in Subsection~\ref{subsec:object_model} and extended its corresponding state vector with four additional entries representing the original delayed output of the system.
A Kalman filter is then used to estimate the new state vector, from which the undelayed object's position is extracted.
% subsection state_estimation (end)

\section{Experiments}
To verify the functionality of the proposed control system, we carried out two experiments.
One for a millimeter-scale spherical object (ball) floating in a shallow pool of water and the other for the same ball on a solid surface.
The parameters of the platform and the control system are summarized in Table~\ref{tb:demmodel_params}.
Videos from the experiments are available at \url{https://youtu.be/BdDq_jRSNYA}.

% Old video (without reference trajecotries) - https://youtu.be/Ntwl7yTYtKg

In the experiments, we manipulate polypropylene balls.
The main reason for this choice of material is its density; the balls made of this plastic float on a water surface, are light enough to be affected by the emitted pressure field and at the same time heavy enough not to be ``blown'' away by it (like, for instance, styrofoam balls we also tested).

\begin{table}[t]
\begin{center}
\captionsetup{width=.9\columnwidth}
\caption{Parameters of the experiments for the floating ball (FB) and the ball on the solid surface (BS).}
\label{tb:demmodel_params}
\begin{tabular}{lccc}
Description &\!\!Notation\!\!& Value & Unit \\\hline
Transducer radius & $r$ & $5$ & \si[per-mode=symbol]{\milli\meter}  \\
Transducer power & $A$ & $6.8$ & \si[per-mode=symbol]{\pascal\meter}  \\
Wavenumber & $k$ & $732.7$ & \si[per-mode=symbol]{\per\meter}  \\
Ball radius (FB and BS) & - & $4$ & \si[per-mode=symbol]{\milli\meter}  \\
\makecell[l]{Transducer to manipulation \\ area vertical distance} & - & $65$ & \si[per-mode=symbol]{\milli\meter}  \\
Pressure point distance & $R$ & $6$ & \si[per-mode=symbol]{\milli\meter}  \\
Pressure to force (FB) & $c_\mathrm{p}$ & $7.65\times 10^{-9}$ & \si[per-mode=symbol]{\newton\per\pascal}  \\
Friction coefficient (FB) & $c_\mathrm{f}$ & $4.44\times 10^{-4}$ & \si[per-mode=symbol]{\newton\second\per\meter}  \\
Pressure Offset (FB) & $P_\mathrm{off}$ & $709.1$ & \si[per-mode=symbol]{\pascal}  \\
Manipulation area radius (FB) & - & $21$ & \si[per-mode=symbol]{\milli\metre}  \\
Pressure to force (BS) & $c_\mathrm{p}$ & $2.64\times 10^{-8}$ & \si[per-mode=symbol]{\newton\per\pascal}  \\
Friction coefficient (BS) & $c_\mathrm{f}$ & $3.18\times 10^{-4}$ & \si[per-mode=symbol]{\newton\second\per\meter}  \\
Pressure Offset (BS) & $P_\mathrm{off}$ & $0$ & \si[per-mode=symbol]{\pascal}  \\
Manipulation area radius (BS) & - & $10$ & \si[per-mode=symbol]{\milli\metre}  \\
\hline
\end{tabular}
\end{center}
\end{table}

\begin{figure*}[t]
    \begin{subfigure}{1\textwidth}
        \includegraphics[width=\textwidth]{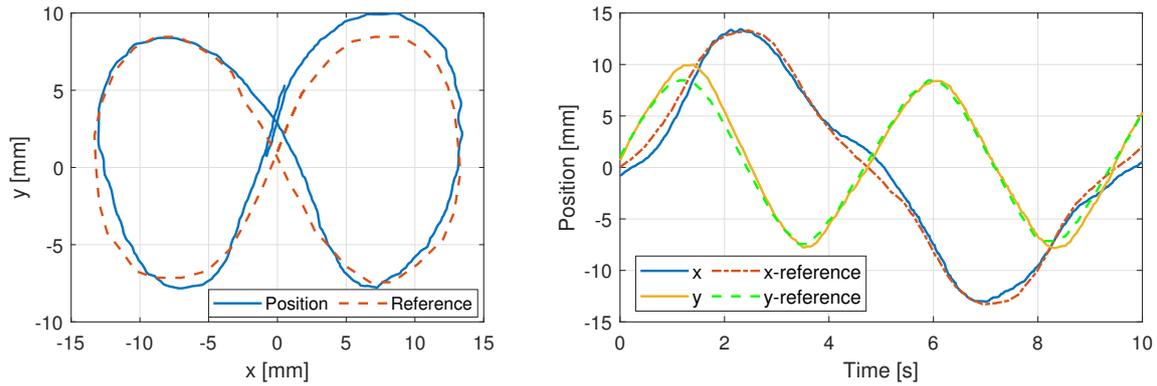}
        \caption{Floating Ball}
        \label{fig:eight}
    \end{subfigure}
    \begin{subfigure}{1\textwidth}
        \includegraphics[width=\textwidth]{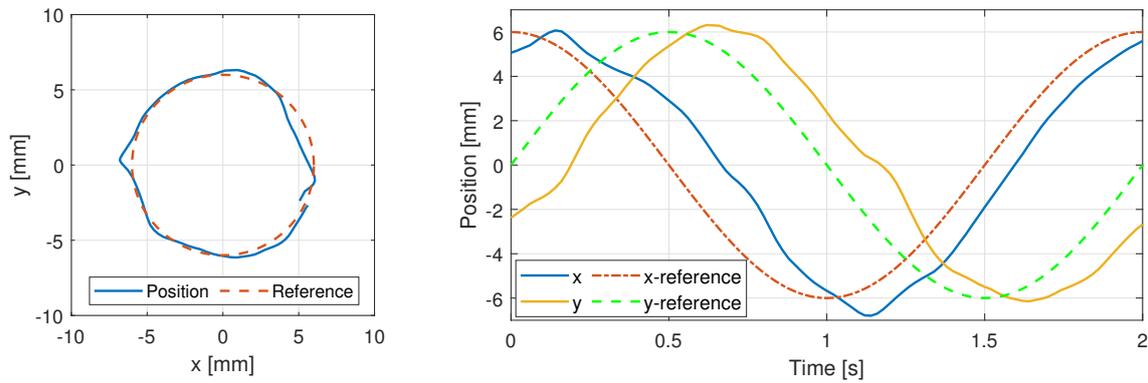}
        \caption{Ball on solid surface}
        \label{fig:solid_exp}
    \end{subfigure}
    \caption{Measured trajectories from the experiments.}
    \label{fig:exps}
    \vspace{-5pt}
\end{figure*}

\subsection{Floating ball}
In the experiment with a floating ball, we have successfully stabilized the particle within a manipulation area with a radius of \SI{21}{\milli\metre}.
Moreover, we were able to track a moving reference with speeds up to \SI{10}{\milli\metre\per\second}.
Example of such tracking is shown in Fig.~\ref{fig:eight} with an eight-figure shaped trajectory.
Even though we are able to generate acoustic pressure with amplitudes up to \SI{3300}{\pascal}, we limit the maximum value to \SI{2500}{\pascal}, as higher-amplitude pressure breaks the surface tension of water, creating bubbles.
These bubbles cause random disturbance to the motion of the object, and thus are not desirable.

% \begin{figure}[h]
% 	\centering
% 	\includegraphics[width = .45 \textwidth]{figures/ref_jump.eps}
% 	\caption{Response of the PID controller to a jump of reference}
% 	\label{fig:ref_jump}
% \end{figure}

% \begin{figure}[h]
% 	\centering
% 	\includegraphics[width = .45 \textwidth]{figures/DB.eps}
% 	\caption{Tracking a moving reference with the floating ball.}
% 	\label{fig:eight}
% \end{figure}

\subsection{Ball on a solid surface}
We managed to stabilize the ball in chosen positions and to follow simple trajectories as it is shown in Fig.~\ref{fig:solid_exp}.
Due to the higher static friction and the faster dynamics, the requirements for the position control of a ball on a solid surface are higher compared to the floating ball.
To steer the ball, we need to generate higher acoustic pressure.
As the maximum achievable pressure decreases with increasing distance from the center, the manipulation area is smaller, having a radius of \SI{10}{\milli\metre}.
We still limit the maximum acoustic pressure to \SI{2700}{\pascal}, because it is reasonable to reduce the acceleration of the ball, and higher values are not achievable evenly in a sufficiently large manipulation area.
% \begin{figure}[h]
% 	\centering
% 	\includegraphics[width = .45 \textwidth]{figures/solid_example1.eps}
% 	\caption{An example of solid surface experiment, see script {\it solidSurfacePlot.m}}
% 	\label{fig:solid_exp1}
% \end{figure}

% \begin{figure}[h]
% 	\centering
% 	\includegraphics[width = .45 \textwidth]{figures/BS.eps}
% 	\caption{An example of solid surface experiment, see script {\it solidSurfacePlot.m}}
% 	\label{fig:solid_exp}
% \end{figure}

\section{Conclusion}
In this paper we presented a novel experimental platform for a non-contact planar manipulation by shaping a pressure (hence force) field generated by an array of ultrasonic transducers.
The manipulated object of a few millimeters in diameter is moved around over a planar surface by generating a high-pressure focal point in its vicinity.
The position and the value of this local maximum of pressure are computed by solving a nonlinear optimization problem in every sampling period of a discrete-time feedback position regulator.
The outcome of such optimization is a set of phase shifts for the voltages applied to the individual transducers in the array.
The position of the object is estimated from the images captured by a digital camera.

The proposed control system was verified by two documented experiments.
Furthermore, the platform was intentionally designed and built using affordable and widely popular components.
Taking into consideration that we made all the relevant technical information available to the public for free reuse, the platform has a potential to become a testbed for algorithms for non-contact planar manipulation.

Even though we demonstrated accurate manipulation with only a single object, the platform can be extended to simultaneous and independent manipulation with several objects.
This would call for expanding the array of transducers (say, to $16\times 16$) and assigning more pressure points in the related optimization problem.
Larger actuator array will also increase the relatively small manipulation area (currently just a few \si{\centi\metre\squared}).

\begin{ack}
This work was supported by the Grant Agency of the Czech Technical University in Prague, grant No. SGS16/232/OHK3/3T/13.
\end{ack}

\bibliography{ultrasonic}             % bib file to produce the bibliography

\appendix

\end{document}